\documentclass[pdflatex,sn-mathphys-num]{sn-jnl}

\usepackage{array}
\newcolumntype{C}[1]{>{\centering\arraybackslash}p{#1}}
\usepackage[utf8]{inputenc}
\usepackage{graphicx}%
\usepackage{multirow}%
\usepackage{amsmath,amssymb,amsfonts}%
\usepackage{amsthm}%
\usepackage{mathrsfs}%
\usepackage[title]{appendix}%
\usepackage{xcolor}%
\usepackage{textcomp}%
\usepackage{manyfoot}%
\usepackage{placeins}
\usepackage{float}
\usepackage{booktabs}%
\usepackage{algorithm}%
\usepackage{algorithmicx}%
\usepackage{algpseudocode}%
\usepackage{listings}%
\usepackage{hyperref}
\usepackage{url}

\theoremstyle{thmstyleone}%
\theoremstyle{thmstyletwo}%

\theoremstyle{thmstylethree}%

\newcommand{\comment}[1]{}

\begin{document}

\title[Article Title]{Clinician-Directed Large Language Model Software Generation for Therapeutic Interventions in Physical Rehabilitation}


\author*[1]{\fnm{Edward} \sur{Kim}}\email{ek65@eecs.berkeley.edu}
\equalcont{These authors contributed equally to this work.}

\author[1]{\fnm{Yuri} \sur{Cho}}
\equalcont{These authors contributed equally to this work.}

\author[2]{\fnm{José Eduardo E.} \sur{Lima}}
\author[2]{\fnm{Julie} \sur{Muccini}}
\author[2]{\fnm{Jenelle} \sur{Jindal}}

\author[3]{\fnm{Alison} \sur{Scheid}}

\author[1]{\fnm{Erik} \sur{Nelson}}
\author[1]{\fnm{Seong Hyun} \sur{Park}}
\author[1]{\fnm{Yuchen} \sur{Zeng}}
\author[1]{\fnm{Alton} \sur{Sturgis}}
\author[1]{\fnm{Caesar} \sur{Li}}
\author[1]{\fnm{Jackie} \sur{Dai}}
\author[1]{\fnm{Sun Min} \sur{Kim}}
\author[1]{\fnm{Yash} \sur{Prakash}}
\author[1]{\fnm{Liwen} \sur{Sun}}
\author[1]{\fnm{Isabella} \sur{Hu}}
\author[1]{\fnm{Hongxuan} \sur{Wu}}
\author[1]{\fnm{Daniel} \sur{He}}
\author[1]{\fnm{Wiktor} \sur{Rajca}}

\author[3]{\fnm{Cathra} \sur{Halabi}}
\author[2]{\fnm{Maarten} \sur{Lansberg}}
\author[1]{\fnm{Bjoern} \sur{Hartmann}}
\author[1]{\fnm{Sanjit A.} \sur{Seshia}}

\affil*[1]{\orgdiv{Electrical Engineering and Computer Sciences}, \orgname{University of California, Berkeley}, \orgaddress{\state{CA}, \country{USA}}}

\affil[2]{\orgdiv{Neurology}, \orgname{Stanford University}, \orgaddress{ \state{CA}, \country{USA}}}

\affil[3]{\orgdiv{Neurology}, \orgname{University of California, San Francisco}, \orgaddress{ \state{CA}, \country{USA}}}


\abstract{Digital health interventions are increasingly used in physical and occupational therapy to deliver home exercise programs via sensor-equipped devices such as smartphones, enabling remote monitoring of adherence and performance. In the current digital intervention paradigm, however, exercise software is typically programmed \textit{before clinical encounters} as libraries of pre-defined modules aimed at broad impairment categories. At the point of care, clinicians can only select from this library and adjust a narrow set of parameters such as duration and repetitions. Patient-specific needs that emerge during encounters—such as distinct movement limitations, personal goals, or home and work constraints—are rarely reflected in the software, limiting personalization and contributing to lower adherence and reduced therapeutic benefit. In this study, we propose a digital intervention paradigm in which large language models (LLMs) act as constrained translators that convert clinicians’ exercise prescriptions into intervention software. Clinicians preserve their role as clinical decision-makers: they design exercises during the encounter, tailored to each patient’s impairments, goals, and environment, and the LLM generates software to match these prescriptions. We conducted a prospective single-arm feasibility study with 20 licensed physical and occupational therapists and a standardized patient. Clinicians created 40 individualized upper extremity exercise programs (398 total instructions), which were automatically translated into executable software. A standardized patient was used because the safety of LLM-generated intervention software had not previously been evaluated. We benchmarked each therapist-designed prescription against a representative template-based DHI to estimate how many could be delivered under the status quo paradigm. The proposed paradigm yielded a 45\% increase in the proportion of personalized prescriptions that could be implemented as software compared with the status quo (100\% vs 55\%; p \textless  0.01), with unanimous agreement among therapists that it was easy to use. The LLM-generated software correctly delivered 99.7\% (397/398) of instructions as prescribed and monitored performance with 88.4\% (95\% CI, 0.843–0.915) accuracy. Overall, 90\% (18/20) of therapists judged the system safe for patient interaction, and 75\% expressed willingness to adopt it in their clinical practice. To our knowledge, this is the first prospective evaluation of clinician-directed intervention software generation with an LLM in healthcare, demonstrating feasibility and motivating larger trials to assess clinical effectiveness and safety in real-patient populations.}


\maketitle

\section{Introduction}\label{sec:intro}

Digital health interventions (DHIs)~\cite{Wienert2022,Tan2024} in physical and occupational therapy increasingly take the form of clinician-prescribed intervention software deployed to a patient’s smartphone or connected wearable device~\cite{Zangger2023,Krohn2024}. Such software delivers step-by-step exercise instructions and can objectively capture adherence and performance using on-device sensors (e.g., camera-based motion tracking on phones or inertial sensing in wearables). As a result, it can quantify whether exercises were completed and how they were performed—such as range of motion, repetition count, tempo, and posture—and return concise summaries to clinicians for timely adjustment of therapy (e.g., modifying difficulty or introducing new exercises). This capability represents a practical advance over standard paper exercise worksheets, which cannot capture how patients perform prescribed exercises at home between visits.

Despite these gains, the prevailing digital prescription paradigm with monitoring remains rigid in its capacity to personalize therapy. In current DHI platforms, \textit{prior to clinical encounters}, therapeutic exercises are encoded as a library of parametrized intervention software, targeting broad populations with common impairments. Consequently, patient-specific needs identifiable only during encounters (for example, distinct deficits, personal goals, or home and work environments) are not fully reflected in the software. Therapists can personalize therapy in limited ways by selecting the most relevant subset of software from the provided library and adjusting a small set of parameters, such as difficulty~\cite{gauthier2022vigorous,prieto2024activehip}, repetitions~\cite{cui2023lowback,malliaras2020intel}, and duration~\cite{weise2022backpain}. This rigid paradigm makes it difficult to flexibly modify exercise instructions—for example, to introduce contingencies when a patient cannot perform a prescribed motion or to incorporate patient-specific house layout into activities of daily living. Prior studies have associated limited personalization in DHIs with lower patient adherence and reduced therapeutic effectiveness~\cite{Palazzo2016,Mahmood2023,Ricke2023}. At the same time, simply expanding configurable options to enhance personalization can impose excessive cognitive load on clinicians in routine care~\cite{Borges2023,vanTilburg2024,Pearce2024}, creating a tension between richer tailoring and practical usability.

\begin{figure}
    \centering
    \includegraphics[width=\linewidth]{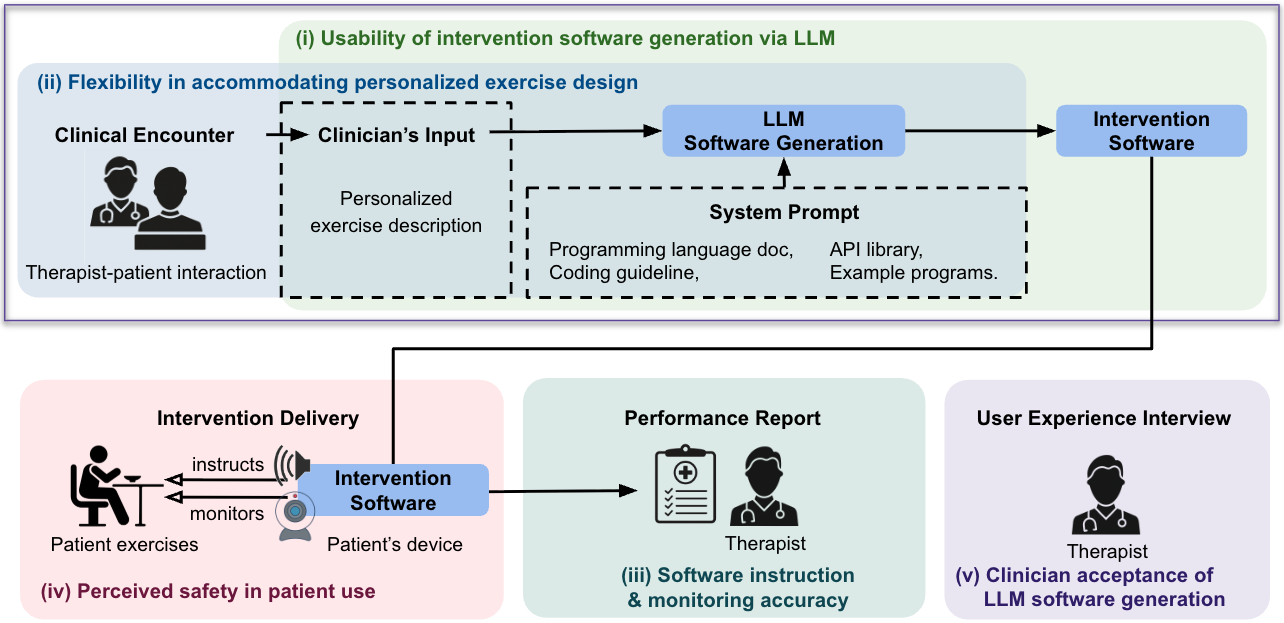}
    \caption{\textbf{Overview of the clinician-directed intervention software generation via LLM and evaluation framework.} Each therapist conducted a therapy session with a standardized patient and designed two tailored exercises, which were translated by the LLM into intervention software. The software was deployed to the standardized patient’s electronic device to instruct the exercises and monitor the patient’s movements, while the therapist observed remotely. Upon completion, the software reported on monitored outcomes back to the therapist. Therapists evaluated the LLM-based digital prescription paradigm across five dimensions: (i) usability of intervention software generation via LLM, (ii) flexibility in accommodating personalized exercise design, (iii) instruction and monitoring accuracy of generated software, (iv) perceived safety in patient use, and (v) clinician acceptance of LLM software generation.}
    \label{fig:overview}
\end{figure}

We posit that advances in large language models (LLMs), particularly their ability to generate software from natural language instructions~\cite{llm_code_gen}, can offer a viable way to resolve this tension. While LLMs have already been used in several healthcare settings~\cite{busch2025llmreview,alkalbani2025llmreview}, their use for generating patient-facing software remains nascent. Early work has applied LLM-based software generation to data retrieval~\cite{Shi2024}, data processing~\cite{Ahn2024}, and visualization~\cite{Liu2025}, but not to the direct creation of clinician-prescribed intervention software. To address the constraints of current DHIs, we propose a digital prescription paradigm in which the LLM functions as a constrained translator rather than a clinical decision-maker. In this paradigm, clinicians design exercises during clinical encounters, tailoring them to each patient’s physical impairments, rehabilitation goals, and home environment. Then, the LLM converts these prescriptions into intervention software without altering their clinical content. By freeing clinicians from pre-defined software templates—while preserving their role as the sole authors of exercise prescriptions—this approach inverts the status quo sequence, generating software \textit{after} the clinician has full knowledge of the patient and allowing the resulting programs to be structurally aligned with individualized patient needs.

We report a prospective, single-arm study involving 20 licensed physical and occupational therapists and a standardized patient to assess the feasibility of clinician-directed, LLM-based software generation from clinicians’ perspectives. We employed a standardized patient because the safety of LLM-generated intervention software had not previously been investigated, and we sought an initial feasibility and safety assessment before enrolling real patients. To our knowledge, the prospective feasibility of clinician-directed LLM software generation has not been evaluated previously in any domain of healthcare. In our study (Fig.~\ref{fig:overview}), each therapist conducted a therapy session with a standardized patient, during which they designed individualized upper extremity exercises comprising 40 exercise programs and 398 total instructions. These exercises were immediately translated into intervention software by the LLM. We then benchmarked the resulting prescriptions against a representative template-based DHI to estimate how many could be delivered under the status quo paradigm and to compare the personalization supported by each approach. Following the session, therapists observed the software instruct the patient, monitor completion of each instruction, and report monitored results back to them for evaluation. Therapists evaluated the LLM-based digital prescription paradigm across five dimensions: (i) usability of intervention software generation via the LLM, (ii) flexibility in accommodating personalized exercise design compared with the existing digital intervention paradigm, (iii) generated software’s accuracy of instruction delivery and monitoring, (iv) perceived safety in patient use, and (v) clinician acceptance of LLM-based software generation.

\section{Results}\label{sec:results}

\subsection{Usability of Intervention Software Generation via LLM}
\textbf{Therapists unanimously found it easy to generate intervention software.}\\
The overview of the evaluation framework is summarized in Fig.~\ref{fig:overview}. 20 licensed physical and occupational therapists each described two upper extremities exercises, which served as input to the LLM. Consistent with standard practice in physical rehabilitation, therapists designed 2 exercises based on their assigned goals: for one goal, each therapist received a prepared worksheet and was asked to edit it to customize the exercise for the patient; for the other goal, therapists designed the exercise from scratch. The LLM subsequently translated them into 40 intervention software comprising a total of 398 individual instructions. Notably, for usability assessment (Fig.~\ref{fig:overview}(i)),  all therapists (20/20) unanimously agreed or strongly agreed (4.5$\pm$0.5) on a 5-point Likert scale that creating personalized software via LLM was easy and intuitive as shown in Fig.~\ref{fig:likert} a) and \ref{fig:likert} b). \\


\vspace{-3mm}
\subsection{Flexibility in Accommodating Personalized Interventions}

\begin{table}
\centering
\caption{\textbf{LLM-enabled personalization categories that are not supported by the current digital intervention paradigm.} The count represents the number of intervention software instances (out of 40 total) that included each personalization category. The percentage indicates the proportion of software instances containing that category.}
\label{tab:infoloss_final}
\begin{tabular}{C{3.5cm} c c C{5.5cm}}
\toprule
\textbf{Category Incompatible} & \textbf{\# of Prescribed} & \textbf{Percentage} & \textbf{Examples} \\
\textbf{with the Status Quo} & \textbf{Exercises} & (out of total 40) & \\
\midrule
Procedural Variation & 15 & 37.5\% & (Shoulder Abduction/Adduction) Substituting a reaching task with a functional task of stacking cubes. \\
\addlinespace
New Equipment Use & 6 & 15.0\% & (Finger Abduction/Adduction) newly introducing a finger web or rubber band for added resistance. \\
\addlinespace
Contingency & 6 & 15.0\% & (Shoulder Flexion) "if the object is too heavy, move your arm out horizontally instead of lifting it high." \\
\addlinespace
Compensatory Strategy Options & 4 & 10.0\% & (Transporting Utensils) Offering biomechanical alternatives: "you may have to rotate your body... or... do shoulder adduction". \\
\addlinespace
Motor Priming & 3 & 7.5\% & (Transporting Beads) Practicing squeezing and relaxing tongs before using them to move beads. \\
\bottomrule
\end{tabular}
\end{table}

\noindent\textbf{Proposed paradigm implemented 45\% more personalized prescriptions than current DHIs.} \\
In the current digital intervention paradigm, physical and occupational therapists personalize care by retrofitting their prescriptions into a pre-programmed library of intervention software written before the clinical encounter. They choose the closest available program and adjust a few parameters (for example, repetitions or difficulty), but cannot change the underlying instructions. In our paradigm, therapists instead write free-text, patient-specific exercise prescriptions after the encounter, and an LLM converts these prescriptions into executable software (Fig.\ref{fig:overview}). We therefore hypothesized that this post-encounter, clinician-directed software generation would allow a larger fraction of personalized prescriptions to be implemented as software than the library-based, retrofit approach used in existing DHIs. To test this, we examined whether each personalized exercise prescribed after the standardized-patient session could be translated into intervention software under (i) our LLM-based paradigm and (ii) a generalized representation of current DHIs (Sec.~\ref{sec:generalized-template}).

Under our proposed paradigm, all 40 personalized prescriptions (100\%) could be translated into intervention software, compared with 22/40 (55\%) under the generalized existing DHI template (Fisher’s exact test, \(p < 0.01\)). All LLM-generated intervention software compiled without syntax errors and were executable. Table~\ref{tab:infoloss_final} summarizes the personalization features that could not be encoded within the existing paradigm (analysis data in Appendix~\ref{supplement:table1}). The most frequent limitation was procedural variation, where therapists substantially changed the exercise procedure—for example, altering the sequence, movement quality, or goals, such as replacing simple shoulder abduction–adduction with a functional cube-transfer task (37.5\%, 15/40). Other common limitations included incorporating new household objects not represented in the template (15\%, 6/40; for example, adding a rubber band to increase resistance during finger abduction–adduction) and contingency rules to cope with difficulties in movement due to patient-specific deficits (15\%, 6/40). Several prescriptions also specified compensatory strategies the template could not encode (10.0\%, 4/40; for example, instructing the patient to rotate the trunk or use shoulder adduction to facilitate task performance).

Together, these findings suggest that clinician-directed software generation substantially expands what digital interventions can express, preserving a larger share of clinically meaningful personalization when translating therapists’ prescriptions into intervention software.


\subsection{Instruction and Monitoring Accuracy of Software}
\begin{figure}
    \centering
    \includegraphics[width=0.95\linewidth]{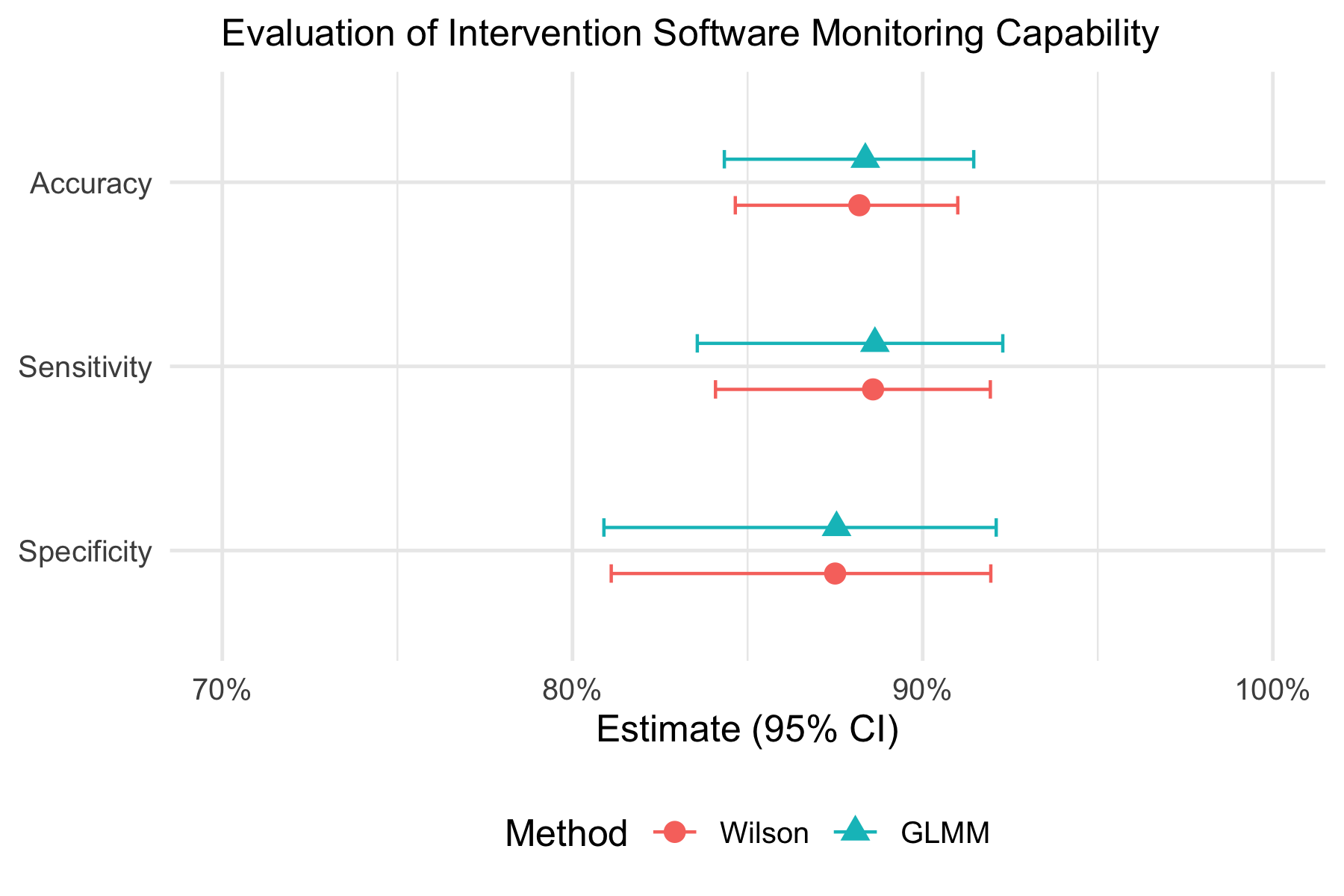}
    \caption{The monitoring accuracy, sensitivity, and specificity of the LLM-generated intervention software with 95\% confidence intervals (CI) were computed using a generalized linear mixed model (GLMM) and a Wilson score test.}
    \label{fig:monitor}
\end{figure}

\noindent\textbf{99.78\% of instructions were correctly delivered by generated software.}\\
After generating intervention software using the LLM, therapists evaluated the fidelity of all 398 software-delivered instructions against their original descriptions, assessing both correctness and completeness (Fig.~\ref{fig:overview}(i)). \textit{Correctness} was defined as the proportion of instructions that exactly matched the content and sequence of the therapist’s descriptions. \textit{Completeness} was defined as the proportion delivered in full, without omissions or extraneous steps (e.g., unnecessary repetition or arbitrary, non-prescribed instructions). Therapists found that 99.78\% (397/398) of instructions were correct and complete. The single instruction error involved an unnecessary repetition that posed no safety risk. There was no generation of arbitrary instructions.\\

\vspace{-2mm}
\noindent\textbf{88.4\% of prescribed instructions were accurately monitored by software.}\\
The LLM-generated intervention software guided the prescribed exercises and monitored the completion of each instruction. Therapists then reviewed the software-generated performance report to assess its monitoring accuracy across all 398 instructions (Fig.~\ref{fig:overview}(iii), and refer to Sec.~\ref{method:monitoring}). The software’s monitoring accuracy, sensitivity, and specificity were 88.4\% (95\% CI, 0.843–0.915), 88.6\% (95\% CI, 0.836–0.923), and 87.5\% (95\% CI, 0.809–0.921), respectively. Confidence intervals were estimated using a generalized linear mixed model (GLMM)~\cite{bolker2009glmmguide} to account for the hierarchical structure of the experiment, in which each exercise goal (e.g., shoulder flexion; exerciseID) was assigned to a therapist (clinicianID), and each therapist created multiple instructions (stepID) for that goal. The GLMM incorporated potential intra-class correlations (ICCs) among exercises, therapists, and instructions when computing accuracy (proportion of correct classifications), sensitivity (proportion of actual successes correctly identified), and specificity (proportion of actual failures correctly identified). To assess the impact of ICCs, we also calculated Wilson score intervals~
\cite{brown2001intervals}, which do not account for correlations. As shown in Fig.~\ref{fig:monitor}, the similarity of the confidence intervals between the two methods indicates that the effects of correlations were minimal. The observed monitoring inaccuracies arose from two sources: hallucinated conditions in the LLM-generated software and errors in visual movement detection. Upon manual inspection, we identified hallucinations in the monitoring logic for 2.5\% (10/398) of instructions, where the generated code checked for conditions that were not specified in the therapist’s prescribed exercise instructions. The remaining inaccuracies predominantly reflected limitations of the computer vision models used to interpret the patient’s movements, leading to incorrect classifications of whether exercises were performed as instructed.

\begin{figure}
    \centering
    \includegraphics[width=0.85\linewidth]{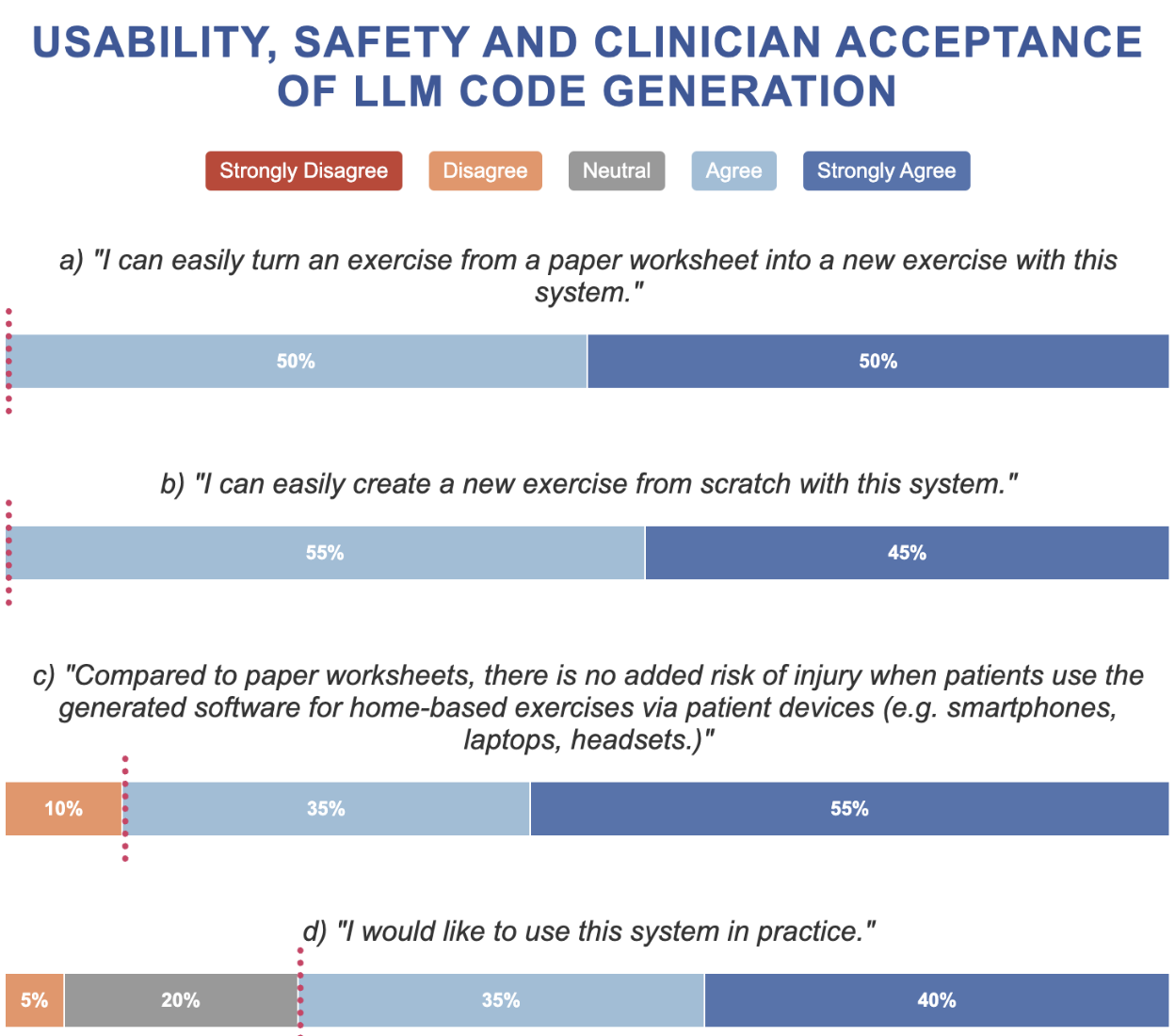}
    \caption{Responses to the 5-point Likert scale questions from the user experience interview on  usability, safety, and clinician acceptance of LLM software generation. Vertical red dotted lines indicate the boundary between scores of 4 (``Agree'') and below. In the survey statements, the term ``system'' referred to the LLM  software generation for ease of understanding by therapists.}
    \label{fig:likert}
\end{figure}

\subsection{Perceived Safety of Intervention Software in Patient Use}
\textbf{90\% of therapists perceived software–patient interactions to be safe.} \\
As shown in Fig.~\ref{fig:overview}(iv), therapists observed the LLM-generated intervention software delivering and monitoring each instruction to assess the safety of its interactions with patients. Overall, 90\% (18/20) of therapists reported that exercising with the software on patients’ personal electronic devices posed no greater risk of injury than using traditional paper worksheets, which remain the most common format for home exercise prescription in standard practice (Fig.~\ref{fig:likert}c). The remaining 10\% (2/20) disagreed, primarily due to concerns about pacing discrepancies—instances in which the software advanced too quickly or too slowly between instructions.

In this context, pacing refers to the software’s ability to wait for patients to complete each instruction before progressing, based on its monitoring of movement. During the interaction, therapists evaluated pacing for each instruction and flagged those they judged inadequate. Overall, 92.8\% (369/398) of instructions were rated as appropriately paced. Inadequate pacing arose from hallucinations in the generated monitoring logic and inaccuracies in the computer vision models. False negatives (i.e., the software failed to detect a completed movement) led to unnecessary delays, whereas false positives (i.e., incomplete movements incorrectly judged as complete) caused the software to advance prematurely.

Therapists noted that such pacing discrepancies, whether too slow or too fast, could lead to unnecessary exertion or discomfort. For example, if the software delays progression while verifying completion of an instruction such as “hold an object in the air,” a patient may maintain the position longer than clinically intended, increasing the risk of fatigue or muscle strain.

\subsection{Clinician Acceptance of LLM software Generation}
\textbf{75\% of therapists expressed willingness to adopt our clinician-generated intervention software paradigm in their clinical practice.}\\
When asked about their willingness to adopt LLM software generation in clinical practice (Fig.~\ref{fig:overview}(v)), 75\% (15/20) of therapists agreed or strongly agreed as shown in Fig.~\ref{fig:likert}d). Many highlighted ease of use and time efficiency, noting that real-time generation of intervention software during clinical encounters allowed them to continue interacting with the patient. A therapist explained, \textit{“It saves time and effort. . . because I could just talk to the patient normally, I’m doing my treatment. . . it’s not eating into productivity time. And then at the same time, it’s creating the home exercise program (software) for me afterwards.”} Others valued the flexibility for generating personalized software, particularly in situations where exercise worksheets or software were not available: \textit{``A lot of the time, for the type of therapy I do, there are no pre-existing handouts. . . So I’m having to kind of create things a lot... I like the fact that this can be customized quickly and easily.}''

Among therapists who were hesitant to adopt, the most frequent feedback concerned the monitoring function of the intervention software. Some wanted greater accuracy: \textit{``They didn’t get all the success or failures correct. . . I don’t know how much I would trust it to be correct when they’re doing it at home when I can’t look at it.''} Others cited the the binary (complete or incomplete) classification did not capture movement quality or compensatory movements, and recommended adding granularity in performance reports about how patients failed, as well as what compensatory strategies were used during successful completions. There were also comments that multi-task instructions required more context-sensitive reporting than a simple pass/fail label: \textit{``If your instruction step is asking multiple things to do and then they fail. . . I would just want more info.''}

\section{Discussion}\label{sec:discussion}
We propose a novel digital intervention paradigm in which clinicians use LLMs to translate their exercise prescriptions into intervention software that can remotely instruct and monitor patients’ exercises. In our study, the generated software was executed on a standardized patient’s device to provide step-by-step guidance, monitor performance, and return monitored outcomes to clinicians for review. Our findings demonstrate that clinician-driven intervention software generation with LLM is feasible and acceptable to therapists in a standardized-patient setting, motivating larger trials to assess clinical effectiveness and safety in real-patient populations.

The proposed digital intervention paradigm enabled a substantially larger fraction of personalized prescriptions to be delivered as software than the existing template-based paradigm, with a 45\% increase in the proportion of prescribed exercises that could be implemented (100\% vs.\ 55\%; p \textless 0.01). This improvement is an expected consequence of how personalization occurs in each paradigm. In our paradigm, therapists author software \textit{after} the clinical encounter, once they have full knowledge of the patient’s impairments, environment, and goals, and use the LLM to generate code that directly reflects their prescriptions. By contrast, in the status quo, intervention software is implemented \textit{before} the encounter, and therapists must retrofit their prescriptions into a fixed library of software, typically by selecting pre-existing software templates and adjusting a narrow set of parameters (for example, repetitions or duration) while the underlying exercise instructions remain fixed. As a result, many clinically important nuances—such as patient-specific household objects used in activities of daily living or contingency instructions when patients struggle with particular motions—cannot be represented, making the observed 45\% gain in deliverable personalization both substantial and anticipated.

Within this paradigm, the LLM-generated intervention software achieved 99.7\% accuracy in delivering exercise instructions (397/398 instructions delivered as prescribed). This near-perfect accuracy is expected: the LLM is not designing exercises or making clinical decisions, but simply transcribing therapists’ prescriptions into executable software. As in standard practice, therapists reviewed the patient chart (here, a clinical vignette summarizing physical deficits in Sec.~\ref{sec:clinical_vignette}), interacted with the standardized patient to elicit needs, and prescribed personalized exercises. The LLM then functioned solely as a constrained translator. As specified in the system prompt (Fig.\ref{fig:overview}), it was instructed to deliver exercise instructions \textit{verbatim as prescribed}, allowing only minor clarifications (for example, resolving ambiguous pronouns such as “this” or “that” based on prior context). In this setting, the task primarily leverages the LLM’s strength in converting text descriptions into structured logic for software execution\cite{llm_code_gen}, rather than requiring complex reasoning. Our use of LLMs is therefore distinct from prior applications that rely on richer clinical reasoning—such as diagnosis\cite{liu2025generalist}, treatment planning~\cite{sandmann2025deepseek}, or risk management~\cite{pais2024medic}. Therapists’ feedback further indicated that this constrained, code-generation role of LLM could be easily integrated into their existing clinical workflow and, in some cases, perceived as enhancing productivity.

In contrast, monitoring whether patients follow exercise instructions correctly is substantially more challenging. It requires the system to interpret each instruction’s intent and relate it to patient's movement data from computer vision models to decide whether the instruction has been completed. Implementing this monitoring logic demands both semantic understanding of the prescribed action and robust handling of noisy, real-world movement data, making it a harder problem than textual transcription. In this setting, we observed monitoring inaccuracies arising from hallucinations in the generated logic and limitations of the computer vision models. Existing DHIs typically rely on libraries that are manually implemented and tested by software engineers, which may, in principle, support higher monitoring reliability than automatically generated programs; our current implementation did not include a manual validation step. Despite these challenges, we observed an 88.4\% monitoring accuracy (352/398 instructions; 95\% CI, 0.84–0.92), which we view as a reasonable preliminary outcome that motivates further investigation of LLM-based software generation for digital interventions. The commercial LLM and computer vision models used here were not fine-tuned for this rehabilitation context; fine-tuning on domain-specific data and incorporating more robust pose-estimation and activity-recognition pipelines are likely to further improve monitoring performance as underlying models advance.

Regarding patient safety, 90\% (18/20) of therapists judged the LLM-generated intervention software to be safe in this standardized-patient setting. Because the safety of such software had not previously been evaluated, we deliberately used a standardized patient rather than real patients in this first feasibility study. This initial safety signal is encouraging but preliminary and requires confirmation in trials with real patients and longer-term use. Therapists observed the software articulate each prescribed instruction and wait until monitoring confirmed completion (or time-to-completion expired) before proceeding, enabling adaptive pacing aligned with patient performance and making sessions more interactive than reading step-by-step instructions on a screen. Most therapists considered the pacing adequate; however, two raised safety concerns about premature or delayed instruction transitions that could pose risks—for example, delayed detection of completing a task such as “hold an object in the air” could lead to fatigue or muscle strain. Improving monitoring accuracy through targeted fine-tuning may reduce premature transitions caused by false positives, while delayed transitions may be mitigated by selecting sensor-processing models—exposed to the LLM as APIs in the system prompt (Fig.~\ref{fig:overview})—with lower latency.

To broadly support physical and occupational therapy with the proposed LLM-based paradigm, our study found that iterative co-design of the system prompt (Fig.~\ref{fig:overview}) between software developers and clinicians was essential. The prompt specifies guidelines for the LLM to generate intervention software, detailing desired functionalities and guardrails aligned with clinical needs. Because authoring such prompts currently requires programming expertise, AI researchers or software engineers remain necessary contributors to developing this class of tools, as with other LLM-based clinical tools. In this study, however, the prompt only partially met therapist expectations, tempering adoption. Some therapists sought finer control over instruction pacing, while others expected monitoring beyond binary labels (complete vs.\ incomplete). Although 75\% (15/20) of therapists indicated willingness to adopt the LLM software-generation tool, with unanimous consensus (20/20) on its ease of use, these unmet needs contributed to hesitancy, underscoring the importance of early needs-finding and iterative prompt refinement to ensure alignment with clinical practice.

Our findings may have implications for other clinical domains that use DHIs. DHIs in mental health~\cite{personalization-ibis-review}, chronic disease management~\cite{ai-diabetes-management}, pain management~\cite{personalizing-pain-management}, and speech–language pathology~\cite{digital-aphasia-therapy} face similar constraints on personalization. As an illustrative (untested) example, a study of exposure and response therapy for obsessive–compulsive disorder (OCD)~\cite{Lenhard2017} did not permit clinicians to encode new task-specific contingencies (for example, “proceed to Step 3 only if hands remain unwashed and visible for 10 minutes”) beyond predefined instructions. In contrast, our approach allows clinicians to specify such contingencies directly in the intervention software (Table~\ref{tab:infoloss_final}). Prospective studies are needed to determine whether the benefits of LLM-based software generation observed here extend to other areas of healthcare.

This study has several limitations. First, we evaluated feasibility solely from clinicians’ perspectives using a standardized patient, without input from real patients. Second, although we hypothesized that enhanced personalization would improve treatment relevance and effectiveness—consistent with prior literature~\cite{golas2021predictive,hwang2023personalization,yardley2015person}—this requires confirmation in longitudinal prospective trials. Third, the study focused on a single patient profile and a small therapist cohort; broader inclusion of diverse conditions and a larger clinician sample will be needed to assess whether LLMs can correctly translate heterogeneous treatment strategies into software with appropriate monitoring logic. Finally, in the absence of benchmark data, we conducted a post hoc comparison against a simulated external benchmark (a generalized software template). Demonstrating superiority in personalization will require a prospective, two-arm trial directly comparing LLM-generated interventions with existing DHI platforms.

In summary, we introduce a novel digital prescription paradigm in which clinicians use an LLM as a constrained translator to generate intervention software \textit{after clinical encounters}, once individual patients’ needs are fully understood. This reverses the prevailing approach, in which clinicians must retrofit prescriptions into pre-programmed software libraries authored \textit{before} encounters. In a prospective feasibility study, this paradigm enabled near-complete translation of personalized prescriptions into software, high accuracy of instruction delivery, and acceptable preliminary monitoring performance, with most therapists judging the system safe and expressing willingness to adopt it in practice. To our knowledge, this is the first prospective evaluation of clinician-directed, LLM-based software generation in healthcare. These findings position LLMs as a practical tool for scaling personalized DHIs and motivate longitudinal prospective trials with real-patient populations.

\section{Methodology}
\subsection{Participants}
We recruited 20 physical and occupational therapists through emails to rehabilitation centers or clinics within the United States. Inclusion criteria were that participants be registered and licensed physical or occupational therapists in the United States and have practiced for at least two years. Written informed consent was obtained before enrollment. This study was reviewed and approved by the institutional review board (UC Berkeley IRB protocol ID: 2025-01-18162). The study lasted for 1 hour per therapist. Each therapist received USD 50 in compensation.

\subsection{Standardized Patient Profile}\label{sec:clinical_vignette}
A standardized patient was employed because the safety of LLM-generated intervention software had not been previously evaluated. The patient presented with right upper extremity weakness characterized as muscle strength graded 3 out of 5 on manual muscle testing, following the Medical Research Council (MRC) scale~\cite{PaternostroSluga2008} for the shoulder, elbow, wrist, and fingers. Strength was preserved to the extent that the patient could move against gravity, but not against resistance. Active right elbow range of motion was limited to 80–120° by a hinged elbow brace with adjustable ROM settings. Cognitive, visual, sensory, and auditory functions were intact. These impairments were simulated using an orthotic splint designed to replicate nonspecific movement limitations commonly observed in outpatient rehabilitation. The clinical vignette was developed in collaboration with licensed therapists and neurologists on the study team to ensure plausibility without attributing the presentation to a specific diagnosis. This approach allowed for consistent presentation across sessions while enabling therapists to tailor exercises to realistic motor constraints without biases related to specific pathologies.

\begin{figure}
    \centering
    \includegraphics[width=0.8\linewidth]{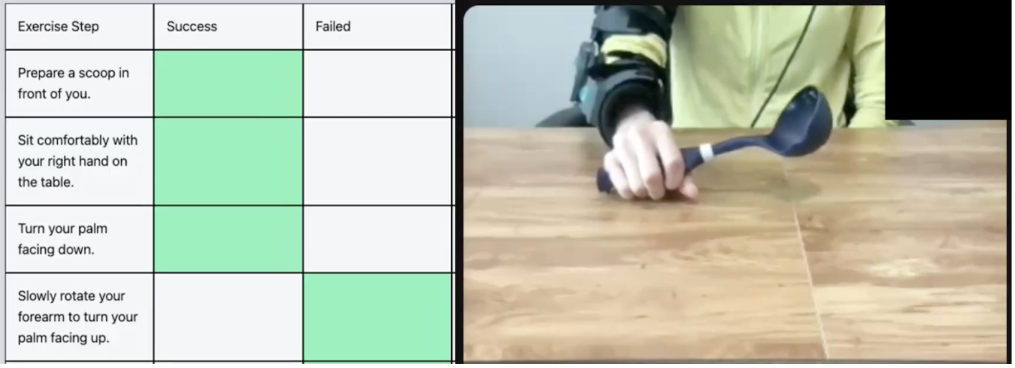}
    \caption{Webportal for clinicians to evaluate monitoring accuracy. On the left, the interface displays a pre-labeled table of completion outcomes, with therapists marking each instruction as success (complete) or failure (incomplete), highlighted in green. On the right, a video panel shows the standardized patient conducting the exercises while following the software’s instructions.}
    \label{fig:evaluation-interface}
\end{figure}

\subsection{LLM Software Generation}\label{method:llm-code-gen}
As shown in Fig.~\ref{fig:overview}, we provided the following system prompt to the LLM in addition to the therapist’s exercise descriptions: (i) documentation on syntax and semantics of the programming language, (ii) an API library, (iii) a coding guideline for software generation, and (iv) example programs pairing sample exercise prescriptions with corresponding exemplary software. The system prompt used for the study can be found in Appendix~\ref{appendix:system_prompt}.\\
\textbf{Programming Language Documentation.} We utilized Scenic~\cite{scenic}, a domain-specific language optimized for modeling and generating physical scenarios; however, this framework is adaptable to any programming language suitable for the targeted electronic devices. \\
\textbf{API Library.} The library of APIs provided function calls to onboard sensors (e.g., camera, inertial measurement unit) and actuators (e.g., speaker) as well as external computation processes (e.g., camera visual data processing for body pose estimation).\\
\textbf{Coding Guideline.} We prompted the LLM to write software in a structured manner such that it iteratively instructs exercises, monitors movements, and logs the patient’s monitored outcomes.\\
\textbf{Example programs.} A curated set of exercise prescriptions and software example pairs was provided to the LLM as a reference on how to write software.

\subsection{Model Selection}
To identify the most suitable model for software generation, we benchmarked four state-of-the-art LLMs available at the time of our experiment (April 2025): Google Gemini 2.5 Pro, OpenAI o4-mini, Claude Sonnet 3.7, and Grok 3.0 Beta. We constructed a dataset comprising 14 distinct instructions and prompted an LLM for software generation, with the inputs aforementioned in Sec.~\ref{method:llm-code-gen}. We manually evaluated the generated software from each model. Grok 3.0-beta demonstrated the highest performance, accurately specifying logic to monitor the completion of all instructions. For this reason, Grok 3.0-beta was used for LLM software generation in our study.

\subsection{Experiment Setup}\label{method:setup}
\textbf{Exercise goals}. We prepared 10 upper extremity exercise goals, each with a corresponding worksheet adapting from Shirley Ryan Ability Lab's upper extremities exercises (e.g., \cite{shirley_ryan}), in consultation with licensed therapists and neurologists on the study team (refer to Appendix~\ref{supplement:exercise_goals}). Each therapist was randomly assigned 2 goals, resulting in 4 therapists per goal (refer to Appendix~\ref{supplement:assigned_exercise_goals}). This study design ensured that all 10 goals were covered and allowed us to examine potential variations in personalization of exercises across therapists. In total, 40 personalized exercise prescriptions are created by therapists (refer to Appendix~\ref{supplement:prescribed_exercises}).

\noindent\textbf{Web portal interface.} We developed a web portal for this study in which therapists conducted live video call sessions with the patient, input exercise descriptions, and evaluated the monitoring capability of the LLM-generated intervention software. Fig. 4 shows the interface used for this evaluation of monitoring accuracy.  

\noindent\textbf{Patient device.} An augmented reality headset (Meta Quest 3) was used. We developed an AR application which executed the intervention software on the headset. The patient wore the headset and followed the verbal instructions from it. The AR application used the speakers as well as the sensors, i.e. RGB cameras and inertial measurement units, to instruct and monitor the patient. 

\subsection{Study Design}\label{sec:study_design}
A prospective single-arm feasibility study was conducted. To mirror clinical practice, therapists reviewed the standardized patient’s clinical vignette and conducted a remote therapy session to design two personalized upper extremities exercises based on their assigned goals: one by revising a prepared worksheet and the other by designing an exercise from scratch. The LLM-generated intervention software then delivered the prescribed exercises and produced a performance report for the therapists.

\subsubsection{Pre-labeled Evaluation for Monitoring Accuracy} \label{method:monitoring}
To evaluate the software’s monitoring capability, therapists completed a pre-labeled assessment. Before observing the software interact with the standardized patient, each therapist viewed a table listing the sequence of instructions the software would deliver (Fig.~\ref{fig:evaluation-interface}, left). For each instruction, therapists indicated in advance whether the patient \textit{should} complete it or leave it incomplete. This design ensured that monitoring accuracy reflected the software’s detection performance rather than the patient’s actual behavior. For example, if the patient completed every step, a system that arbitrarily marked all instructions as complete would appear to be 100\% accurate without actually monitoring. Pre-labeling created a known mix of complete and incomplete instructions, allowing us to compute accuracy, sensitivity, and specificity; overall, 36.3\% of instructions were pre-labeled as incomplete.

\subsubsection{Comparison of Existing and Proposed DHI Paradigms}\label{sec:generalized-template}

Current commercial DHIs for remote exercise instruction typically provide libraries of pre-programmed intervention software that instantiate standard exercise worksheets targeting broad patient populations. These modules encode fixed textual instructions (for example, generic upper extremity strengthening programs) and expose a limited set of parameters—such as number of repetitions, difficulty level, or target side—to tailor treatment to individual patients. In routine use, clinicians choose a module that roughly matches a patient’s impairment and then adjust these parameters. However, they generally cannot modify the underlying instructions to incorporate patient-specific contingencies (for example, alternative steps when a patient cannot perform a motion) or contextual details (for example, household objects, home layout, or work schedule). Because existing systems are heterogeneous and proprietary, there is no standardized benchmark or parameter schema describing how such exercise libraries are structured.

To enable a systematic comparison with this prevailing digital intervention paradigm, we constructed a generalized DHI template based on standardized exercise worksheets and recent digital rehabilitation trials in physical and occupational therapy. We first derived a generalized set of exercise parameters by aggregating those reported in digital rehabilitation trials over the past ten years, including exercise side (left/right/bilateral)~\cite{shebib2019lowback,pak2023shoulderpain}, number of repetitions~\cite{cui2023lowback,malliaras2020intel,mecklenburg2018kneepain}, pose hold time~\cite{weise2022backpain}, and predefined difficulty or goal options (for example, ranges of motion or target locations)~\cite{prieto2024activehip,pak2023shoulderpain,gauthier2022vigorous,bailey2020chronicpain,biebl2021app}. We then instantiated this generalized template using ten exercise worksheets adapted from existing standardized worksheets, each corresponding to a common exercise goal (Sec.~\ref{method:setup}). In this generalized DHI template, the worksheet instructions represent the fixed exercise content, and clinicians are allowed to adjust only the template parameters, mirroring how current DHIs permit parameter tuning without altering the underlying instructions.

After each clinical encounter with the standardized patient, therapists prescribed personalized exercises for their assigned exercise goals (Appendix~\ref{supplement:exercise_goals},\ref{supplement:assigned_exercise_goals}); each goal was assigned to four different therapists to capture variation in personalized design (Sec.~\ref{method:setup}). Our intent was to quantify how these post-encounter, therapist-authored prescriptions deviate from the standard worksheet–based instructions that underpin current DHIs, and to assess whether those deviations could be accommodated by parameter adjustments alone. For each of the 40 personalized prescriptions, we therefore examined whether it could be retrofitted into the generalized DHI template for the corresponding exercise goal by (i) selecting the associated worksheet-based instruction set and (ii) adjusting only the allowed parameters. If a prescription required changing the textual instructions (for example, adding contingencies or patient-specific context) rather than just tuning parameters, we considered it not translatable into intervention software within the current digital intervention paradigm.



\subsection{Data Availability \& Reproducibility}
The study data including exercise prescriptions from therapists, the corresponding generated intervention programs, and the code and prompt for LLM-based intervention software generation can be found in this GitHub repository (\url{https://github.com/ek65/nature_medicine_data_analysis.git}). Also, for reproducibility, the code for executing the generated intervention software on a patient's device can be found in this repository (\url{https://github.com/ek65/nature_medicine_experiment_setup.git}).

\section{Funding}
This work was supported by the UC Noyce Initiative’s Computational Health Program, which funds innovative research at the intersection of computing, digital transformation, medicine, and the health sciences, and supported the collaboration between UC Berkeley and UC San Francisco. Authors affiliated with Stanford University contributed to this work without dedicated funding support.

\section{Conflict of Interest}
All of the authors declared that they have no conflicts of interest.

\section{Author Contributions}
Edward Kim (E.K.) and Yuri Cho (Y.C.) co-led the research. E.K. identified the core problem, co-designed and conducted the study, led the overall technical direction, architected and implemented the LLM-based intervention software system, coordinated system development, and wrote the Results and Methods sections. Y.C. led the clinical study design, formulated the study hypotheses, conducted the study, and drafted the Introduction and Discussion sections. José Lima (J.L.) refined the study design and performed the statistical analyses. Julie Muccini (J.M.) refined the study design, particularly the definition of exercise goals and worksheets, and iteratively tested the system, providing feedback from an occupational therapy perspective. Jenelle Jindal (J.J.) contributed to refining the study design. Allison Scheid (A.Sc.) tested the LLM-based tool and provided feedback from a physical therapy perspective. Erik Nelson (E.N.) designed and implemented the back-end server used in the study. Sun Min Kim (S.K.) implemented the database and cloud-based program synthesis routines. Seong Hyun Park (S.P.) led the design and implementation of the augmented reality (AR)–based patient-facing application and its integration with the tele-rehabilitation platform, enabling remote therapist–patient sessions. Alton Sturgis (A.St.) implemented the video call functionality of the tele-rehabilitation platform in the AR headset. Caesar Li (C.L.) implemented real-time body pose estimation to detect and analyze movements. Jackie Dai (J.D.) developed algorithms to capture images of body movements via AR headset cameras for real-time analysis. Yash Prakash (Y.P.) designed and implemented the multiplayer Unity server that enabled clinician–patient interaction within the tele-rehabilitation platform. Daniel He (D.H.) integrated Scenic with Unity to support instruction and monitoring of patient body movements. Wiktor Rajca (V.R.) developed algorithms to parse movement trajectories into interpretable joint angles for real-time analysis. Yuchen Zeng (Y.Z.) led development of the web portal used by clinicians to conduct sessions and prescribe exercises. Liwen Sun (L.S.) designed and implemented the evaluation schema therapists used to assess patients via the portal. Isabella Hu (I.H.) architected the portal’s video call functionality and workflow for recording and parsing sessions to generate exercise worksheets using LLMs. Hongxuan Wu (H.W.) led the visual design of the portal. Dr. Cathra Halabi (C.H.) refined the study design and provided clinical oversight from a neurology perspective. Dr. Maarten Lansberg (M.L.) refined the study design, contributed to interpretation of results, and advised on statistical analysis. Professors Sanjit Seshia (S.S.) and Bjoern Hartmann (B.H.) advised the project from the computer science side, helped refine the study design, supervised system development, and served as technical co–principal investigators. Dr. Halabi and Dr. Lansberg served as clinical principal investigators. All authors reviewed, edited, and approved the final manuscript.

\backmatter

\begin{appendices}

\section{System Prompt to LLM}\label{appendix:system_prompt}
The system prompt shown in Fig.~\ref{fig:overview} can be found in this github repository (\url{https://github.com/ek65/nature_medicine_data_analysis.git}).

\section{Provided Exercise Instructions}\label{supplement:exercise_goals}
We prepared the following 10 exercise goals and corresponding instructions, adapting the style from \href{https://www.sralab.org/sites/default/files/2017-05/Upper%20Body%20Arm%20strengthening%20exercises%20with%20tabletop%20support.pdf}{Shirley Ryan Ability Lab's upper extremities exercises}.\\

\noindent\textbf{Exercise Goal 1: Shoulder abduction and adduction}
\begin{enumerate}
\item Sit at a table with pink, blue, and yellow post-its.
\item With your left hand, place the pink post-it on the table to the left of you by about 10 inches.
\item Place the blue post-it in the middle.
\item Place the yellow post-its on the right side by about 10 inches.
\item Position your right hand on the blue post-it.
\item Move your hand across the body to touch the pink post-it.
\item Move your hand to the side to touch the yellow post-it.
\item Move your hand across the body to touch the pink post-it.
\item Move your hand to the side to touch the yellow post-it.
\item Return your hand to the blue post-it.
\item Relax your right hand.
\end{enumerate}

\medskip

\noindent\textbf{Exercise Goal 2: Elbow extension using two objects}
\begin{enumerate}
\item Sit at a table with a pink and a yellow post-it in front of you.
\item Rest your right hand on the table.
\item With your left hand, place the pink post-it at the edge of the table away from you.
\item With your left hand, place the yellow post-it at the edge of the table right in front of you.
\item Position your right hand on the yellow post-it.
\item Reach forward with your right hand to touch the pink post-it by extending your elbow.
\item Return your hand to the yellow post-it.
\item Reach forward with your right hand to touch the pink post-it by extending your elbow.
\item Return your hand to the yellow post-it.
\item Rest your hand on the table.
\end{enumerate}

\medskip

\noindent\textbf{Exercise Goal 3: Forearm supination and pronation}
\begin{enumerate}
\item Prepare a scoop in front of you.
\item Sit comfortably with your right hand on the table.
\item Turn your palm facing down.
\item Slowly rotate your forearm to turn your palm facing up.
\item Rotate your forearm again to turn your palm facing down.
\item Grab a scoop with your right hand.
\item Rotate your forearm to turn your palm up while holding the scoop.
\item Rotate your forearm to turn your palm down while holding the scoop.
\item Place the scoop on the table.
\item Rest your right hand.
\end{enumerate}

\medskip

\noindent\textbf{Exercise Goal 4: Thumb opposition}
\begin{enumerate}
\item Sit comfortably with your right hand resting on a table.
\item Open your right hand with fingers relaxed and slightly spread.
\item Touch the tip of your right thumb to the tip of your index finger.
\item Release the thumb and index fingers.
\item Touch the tip of your thumb to the tip of your middle finger.
\item Release the thumb and middle fingers.
\item Touch the tip of your thumb to the tip of your ring finger.
\item Release the thumb and ring fingers.
\item Touch the tip of your thumb to the tip of your pinky.
\item Release the thumb and the pinky.
\item Rest your right hand.
\end{enumerate}

\medskip

\noindent\textbf{Exercise Goal 5: Finger abduction/adduction}
\begin{enumerate}
\item Rest your right hand on the table with your palm facing down.
\item Spread apart your index and middle fingers as much as you can.
\item Bring them back together.
\item Spread apart your middle and ring fingers.
\item Bring them back together.
\item Spread apart your ring and little fingers.
\item Bring them back together.
\item Spread apart your thumb and index fingers.
\item Bring them back together.
\item Rest your right hand.
\end{enumerate}

\medskip

\noindent\textbf{Exercise Goal 6: Move fruits out of a bowl}
\begin{enumerate}
\item Sit at the table with a bowl containing an apple, a lemon, and a banana in front of you.
\item Place your right hand near the bowl.
\item Use your right hand to pick up the apple from the bowl.
\item Place the apple gently on the table.
\item Use your right hand to pick up the lemon from the bowl.
\item Gently place the lemon on the table.
\item Place your right hand on the table to take some rest.
\item Pick up the banana using your right hand.
\item Place the banana on the table.
\item Rest your right arm.
\end{enumerate}

\medskip

\noindent\textbf{Exercise Goal 7: Putting utensils into a container}
\begin{enumerate}
\item Sit at the table with a spoon, a pair of chopsticks, a fork, and a container in front of you.
\item Place your right hand on the table near the utensils.
\item Use your right hand to pick up the spoon.
\item Place the spoon into the container.
\item Rest your hand on the table to take a brief break.
\item Pick up the pair of chopsticks with your right hand.
\item Place the chopsticks into the container.
\item Rest your hand on the table to take a brief break.
\item Pick up the fork with your right hand.
\item Place the fork into the container.
\item Rest your right arm.
\end{enumerate}

\medskip

\noindent\textbf{Exercise Goal 8: Remove coins from a wallet}
\begin{enumerate}
\item Sit at a table with a wallet placed in front of you.
\item Place your right hand near the wallet.
\item Hold the wallet with your left hand.
\item Use your right hand to unzip the wallet slowly.
\item Use your right hand to reach into the coin pocket.
\item Pinch and remove one coin using your fingers.
\item Place the coin on the table.
\item Pinch and remove one more coin using your fingers.
\item Place the coin on the table.
\item Place the wallet on the table.
\item Rest your right arm.
\end{enumerate}

\medskip

\noindent\textbf{Exercise Goal 9: Stack three cubes}
\begin{enumerate}
\item Sit at the table with orange, blue, and green cubes in front of you.
\item Use your right hand to pick up the orange cube.
\item Place the orange cube on the table as the base.
\item Pick up the blue cube with your right hand.
\item Stack the blue cube on top of the orange cube.
\item Pick up the green cube with your right hand.
\item Stack the green cube on top of the blue cube.
\item Grab the green cube with your right hand.
\item Place the green cube on the table.
\item Grab the blue cube with your right hand.
\item Place the blue cube on the table.
\item Rest your right hand.
\end{enumerate}

\medskip

\noindent\textbf{Exercise Goal 10: Transport beads from a bowl to a towel}
\begin{enumerate}
\item Sit at the table with a bowl containing three colored spheres---yellow, red, and blue---, a towel, and a tong in front of you.
\item Pick up the tongs with your right hand.
\item Use the tongs to pick up the yellow bead from the bowl.
\item Move the yellow bead onto the towel and release it gently.
\item Use the tongs to pick up the red bead from the bowl.
\item Move the red bead onto the towel and release it.
\item Pick up the blue bead from the bowl using the tongs.
\item Place the blue sphere on the towel.
\item Place the tongs on the table.
\item Rest your right hand.
\end{enumerate}

\section{Therapists' Exercise Prescriptions}\label{supplement:prescribed_exercises}
This study data can be found in our github repository, (\url{https://github.com/ek65/study_data_analysis}).

\section{Assigned Exercise Goals}\label{supplement:assigned_exercise_goals}
The two exercise IDs (listed in Appendix~\ref{supplement:exercise_goals}) assigned to each therapist in the study are listed below. 

\begin{table}[htbp]
\centering
\caption{Assigned Exercise IDs by Participant 1 through 10}
\begin{tabular}{l*{10}{c}}
\toprule
\textbf{Therapist ID} & \textbf{1} & \textbf{2} & \textbf{3} & \textbf{4} & \textbf{5} & \textbf{6} & \textbf{7} & \textbf{8} & \textbf{9} & \textbf{10} \\
\midrule
1st Assigned Exercise ID & 5 & 1 & 9 & 4 & 10 & 2 & 6 & 7 & 8 & 3 \\
2nd Assigned Exercise ID & 10 & 6 & 3 & 9 & 1 & 7 & 4 & 5 & 2 & 6 \\
\bottomrule
\end{tabular}%
\end{table}

\begin{table}[htbp]
\centering
\caption{Assigned Exercise IDs by Participant 11 through 20}
\begin{tabular}{l*{10}{c}}
\toprule
\textbf{Therapist ID} & \textbf{11} & \textbf{12} & \textbf{13} & \textbf{14} & \textbf{15} & \textbf{16} & \textbf{17} & \textbf{18} & \textbf{19} & \textbf{20} \\
\midrule
1st Assigned Exercise ID & 4 & 1 & 9 & 6 & 3 & 2 & 8 & 7 & 5 & 10 \\
2nd Assigned Exercise ID & 7 & 8 & 5 & 2 & 9 & 10 & 4 & 1 & 8 & 3 \\
\bottomrule
\end{tabular}%
\end{table}

\section{Justification of Table~\ref{tab:infoloss_final}}\label{supplement:table1}

\subsubsection*{\textbf{Procedural Variation}}
\textbf{(Count: 15)} This feature identifies instructions where the core sequence, quality of movement, or therapeutic goal is substantially different from the template.

\begin{itemize}
  \item \textbf{Source: Exercise 1 (Therapist 5)}
    \begin{itemize}
      \item \textbf{Justification:} Deviates from the template’s 2D reaching task by introducing a different 3D functional goal of moving cubes to specific locations.
    \end{itemize}
    \begin{quote}
      8. Move the green cube to the pink post-it note.\\
      9. Move the blue cube to the blue post-it note.\\
      10. Move the orange cube to the yellow post-it note.
    \end{quote}

  \item \textbf{Source: Exercise 1 (Therapist 18)}
    \begin{itemize}
      \item \textbf{Justification:} Introduces a completely different procedure of lining up cubes and then moving them to the edge of the table.
    \end{itemize}
    \begin{quote}
      1. Line up all three cubes on Yuri's left side using the right hand.\\
      3. Bring the orange cube to the edge of the table on the right side.\\
      7. Place the green cube under the blue cube, at the edge of the table of the right side.
    \end{quote}

  \item \textbf{Source: Exercise 2 (Therapist 14)}
    \begin{itemize}
      \item \textbf{Justification:} The procedure is entirely different, changing the goal from a horizontal reach to a vertical lift against gravity using weighted objects.
    \end{itemize}
    \begin{quote}
      2. Lift the green cube up to your shoulder by bending your right elbow.\\
      3. Lower the green cube back down by straightening your elbow as much as you can.
    \end{quote}

  \item \textbf{Source: Exercise 3 (Therapist 3)}
    \begin{itemize}
      \item \textbf{Justification:} The repetition structure is different from the template, including explicit ``break'' and ``repeat'' steps. It also adds a safety contingency.
    \end{itemize}
    \begin{quote}
      5. Repeat the rotation with the scoop once more, going palm down then palm up.\\
      6. Take a break until you can begin again .\\
      7. Repeat the exercise 1 more time.\\
      Stop if there is any pain.
    \end{quote}

  \item \textbf{Source: Exercise 3 (Therapist 15)}
    \begin{itemize}
      \item \textbf{Justification:} Adds a new instruction to lift the hand
    \end{itemize}
    \begin{quote}
      6. Lift your hand from the table, while stabilizing your elbow on the table.
    \end{quote}

  \item \textbf{Source: Exercise 3 (Therapist 20)}
    \begin{itemize}
      \item \textbf{Justification:} Modifies the procedure by adding an isometric hold, a different therapeutic element.
    \end{itemize}
    \begin{quote}
      5. Hold each position for 3 seconds.
    \end{quote}

  \item \textbf{Source: Exercise 5 (Therapist 8)}
    \begin{itemize}
      \item \textbf{Justification:} The procedure is modified with a conditional rule for progression.
    \end{itemize}
    \begin{quote}
      Note: If the exercise is not challenging enough, additional resistance such as finger web or rubber band may be used to make it harder.
    \end{quote}

  \item \textbf{Source: Exercise 5 (Therapist 13)}
    \begin{itemize}
      \item \textbf{Justification:} The sequence is different. The template instructs individualized finger movements, while this instruction prescribes global finger adduction/abduction.
    \end{itemize}
    \begin{quote}
      3. Bring all your fingers together so that there's no space in between your fingers.\\
      4. Spread your fingers out all the way.
    \end{quote}

  \item \textbf{Source: Exercise 6 (Therapist 10)}
    \begin{itemize}
      \item \textbf{Justification:} Deviates from the template by introducing cubes and a different movement pattern focused on lifting against gravity.
    \end{itemize}
    \begin{quote}
      3. Bring all your fingers together so that there's no space in between your fingers.\\
      4. Spread your fingers out all the way.
    \end{quote}

  \item \textbf{Source: Exercise 7 (Therapist 6)}
    \begin{itemize}
      \item \textbf{Justification:} The entire procedure is different, focusing on individual grasp-and-release cycles with adaptive strategies, unlike the template's simple sequence.
    \end{itemize}
    \begin{quote}
      3. Take the lighter weight object, the cube.\\
      4. Pick up the cube and try to lift it as much as you can.
    \end{quote}

  \item \textbf{Source: Exercise 7 (Therapist 11)}
    \begin{itemize}
      \item \textbf{Justification:} The procedure is fundamentally altered by adding a preparatory motor priming phase before the main task, and also specifies how to exactly pick up the spoon which was not specified in the original template instruction.
    \end{itemize}
    \begin{quote}
      1. Perform opposition (touch the pads of your thumb and index finger) between your thumb and index finger.\\
      5. Use your thumb and index finger to pick up the handle of the spoon.
    \end{quote}

  \item \textbf{Source: Exercise 8 (Therapist 12)}
    \begin{itemize}
      \item \textbf{Justification:} The sequence is modified by adding post-task practice steps.
    \end{itemize}
    \begin{quote}
      10. ``Practice picking up the coin from the table and then putting it down.''\\
      12. ``Practice the action of putting the coin back in the wallet, using this as a gripping exercise with your fingers.''
    \end{quote}

  \item \textbf{Source: Exercise 8 (Therapist 19)}
    \begin{itemize}
      \item \textbf{Justification:} Adds a novel functional step (stacking) not present in the template.
    \end{itemize}
    \begin{quote}
      6. Stack this coin on top of the other coin.
    \end{quote}

  \item \textbf{Source: Exercise 9 (Therapist 15)}
    \begin{itemize}
      \item \textbf{Justification:} The procedure is entirely different, involving a different stacking order and adding complex movements like crossing the body.
    \end{itemize}
    \begin{quote}
      6. Stack the green cube on top of the blue cube.\\
      7. Cross your body with your right arm towards the orange cube.
    \end{quote}

  \item \textbf{Source: Exercise 10 (Therapist 16)}
    \begin{itemize}
      \item \textbf{Justification:} The procedure is modified by adding a second phase where the items are transferred back to their starting point.
    \end{itemize}
    \begin{quote}
      6. after all balls are transferred to the towel, transfer them back to the bowl one at a time
    \end{quote}
\end{itemize}

\noindent\rule{\linewidth}{0.4pt}

\subsubsection*{\textbf{Equipment Modification}}
\textbf{(Count: 6)} This feature identifies instructions that use different or additional physical objects compared to the template.

\begin{itemize}
  \item \textbf{Source: Exercise 1 (Therapist 5)}
    \begin{itemize}
      \item \textbf{Justification:} Uses ``cubes'' in addition to the template's post-it notes.
    \end{itemize}
    \begin{quote}
      8. Move the green cube to the pink post-it note.
    \end{quote}

  \item \textbf{Source: Exercise 1 (Therapist 18)}
    \begin{itemize}
      \item \textbf{Justification:} Uses ``cubes'' instead of the template's post-it notes.
    \end{itemize}
    \begin{quote}
      1. Line up all three cubes...
    \end{quote}

  \item \textbf{Source: Exercise 2 (Therapist 14)}
    \begin{itemize}
      \item \textbf{Justification:} Uses ``cubes'' instead of the template's post-it notes.
    \end{itemize}
    \begin{quote}
      1. Grab the green cube with your right hand.
    \end{quote}

  \item \textbf{Source: Exercise 5 (Therapist 8)}
    \begin{itemize}
      \item \textbf{Justification:} Suggests using a ``finger web or rubber band,'' which are not in the template.
    \end{itemize}
    \begin{quote}
      Note: ...additional resistance such as finger web or rubber band may be used...
    \end{quote}

  \item \textbf{Source: Exercise 6 (Therapist 10)}
    \begin{itemize}
      \item \textbf{Justification:} Uses ``cubes'' in addition to the template's fruits.
    \end{itemize}
    \begin{quote}
      3. ``Take the lighter weight object, the cube.''
    \end{quote}

  \item \textbf{Source: Exercise 7 (Therapist 6)}
    \begin{itemize}
      \item \textbf{Justification:} Substitutes the template's spoon and chopsticks with a ``scoop'' and ``tongs.''
    \end{itemize}
    \begin{quote}
      1. ...grasp the of the scoop handle.\\
      8. ...grasp the handle tongs.
    \end{quote}
\end{itemize}

\noindent\rule{\linewidth}{0.4pt}

\subsubsection*{\textbf{Safety \& Conditional Logic}}
\textbf{(Count: 6)} This feature identifies ``if-then'' statements or explicit safety warnings.

\begin{itemize}
  \item \textbf{Source: Exercise 3 (Therapist 3)}
    \begin{itemize}
      \item \textbf{Justification:} Provides an explicit safety contingency based on pain.
    \end{itemize}
    \begin{quote}
      Stop if there is any pain.
    \end{quote}

  \item \textbf{Source: Exercise 5 (Therapist 8)}
    \begin{itemize}
      \item \textbf{Justification:} Provides a conditional rule for progressing the difficulty.
    \end{itemize}
    \begin{quote}
      Note: If the exercise is not challenging enough, additional resistance... may be used...
    \end{quote}

  \item \textbf{Source: Exercise 6 (Therapist 2)}
    \begin{itemize}
      \item \textbf{Justification:} Provides a conditional strategy based on the weight of the object.
    \end{itemize}
    \begin{quote}
      4. ...You can use your left arm to support your right arm if the apple is too heavy.
    \end{quote}

  \item \textbf{Source: Exercise 6 (Therapist 10)}
    \begin{itemize}
      \item \textbf{Justification:} Provides a conditional motor adaptation based on difficulty.
    \end{itemize}
    \begin{quote}
      10. If the object is too heavy, move your arm out horizontally instead of lifting the banana high.
    \end{quote}

  \item \textbf{Source: Exercise 7 (Therapist 6)}
    \begin{itemize}
      \item \textbf{Justification:} Provides a conditional strategy if the object is too heavy.
    \end{itemize}
    \begin{quote}
      9. If the tongs are to heavy to bring over, use your left arm to move the container towards the middle...
    \end{quote}

  \item \textbf{Source: Exercise 7 (Therapist 11)}
    \begin{itemize}
      \item \textbf{Justification:} Provides a conditional grip adaptation based on the weight of the object.
    \end{itemize}
    \begin{quote}
      8. If the fork is a bit heavy, use your middle finger together with your thumb and index finger to form a tripod grasp.
    \end{quote}
\end{itemize}

\noindent\rule{\linewidth}{0.4pt}

\subsubsection*{\textbf{Motor Priming}}
\textbf{(Count: 3)} This feature identifies preparatory exercises embedded within a larger functional task.

\begin{itemize}
  \item \textbf{Source: Exercise 7 (Therapist 11)}
    \begin{itemize}
      \item \textbf{Justification:} Prescribes a full set of thumb opposition exercises before the patient is asked to use that grasp to pick up a utensil.
    \end{itemize}
    \begin{quote}
      1. Perform opposition... between your thumb and index finger.\\
      2. Try the same opposition with your thumb and middle finger.\\
      3. Next, try the opposition with your thumb and ring finger.\\
      4. Attempt the opposition with your thumb and pinkie finger.
    \end{quote}

  \item \textbf{Source: Exercise 8 (Therapist 17)}
    \begin{itemize}
      \item \textbf{Justification:} Includes practice of a component skill after the main task, priming the motor pattern for subsequent task of putting the coin back into the wallet.
    \end{itemize}
    \begin{quote}
      10. ``Practice picking up the coin from the table and then putting it down.''\\
      11. Practice picking up the coin from the table and then putting it down, ensuring to bend your elbow during the process.\\
      12. Practice the action of putting the coin back in the wallet, using this as a gripping exercise with your fingers.
    \end{quote}

  \item \textbf{Source: Exercise 10 (Therapist 1)}
    \begin{itemize}
      \item \textbf{Justification:} Includes an explicit preparatory step to practice the required muscle action.
    \end{itemize}
    \begin{quote}
      2. Squeeze the tongs together and relax it.
    \end{quote}
\end{itemize}

\noindent\rule{\linewidth}{0.4pt}

\subsubsection*{\textbf{Compensatory Strategy Options}}
\textbf{(Count: 4)} This feature identifies instructions that offer alternative ways to perform a movement, often by using other body parts.

\begin{itemize}
  \item \textbf{Source: Exercise 6 (Therapist 2)}
    \begin{itemize}
      \item \textbf{Justification:} Offers the use of the left arm to support the right arm.
    \end{itemize}
    \begin{quote}
      4. ...You can use your left arm to support your right arm if the apple is too heavy.
    \end{quote}

  \item \textbf{Source: Exercise 6 (Therapist 10)}
    \begin{itemize}
      \item \textbf{Justification:} Provides a conditional motor adaptation based on difficulty.
    \end{itemize}
    \begin{quote}
      10. If the object is too heavy, move your arm out horizontally instead of lifting the banana high.
    \end{quote}

  \item \textbf{Source: Exercise 7 (Therapist 6)}
    \begin{itemize}
      \item \textbf{Justification:} Offers alternative body movements (trunk rotation or shoulder adduction) to achieve the goal.
    \end{itemize}
    \begin{quote}
      3. ...you may have to rotate your body towards the container, or you may have to do shoulder adduction to get closer to the container.
    \end{quote}

  \item \textbf{Source: Exercise 7 (Therapist 11)}
    \begin{itemize}
      \item \textbf{Justification:} Provides a conditional grip adaptation based on the weight of the object.
    \end{itemize}
    \begin{quote}
      8. If the fork is a bit heavy, use your middle finger together with your thumb and index finger to form a tripod grasp.
    \end{quote}
\end{itemize}





\end{appendices}


\bibliography{sn-bibliography}

\clearpage
\end{document}